\documentclass[fleqn]{SelfArx}

\definecolor{color1}{RGB}{0,0,90} % Color of the article title and sections
\definecolor{color2}{RGB}{0,20,20} % Color of the boxes behind the abstract and headings

\usepackage{hyperref}
\hypersetup{hidelinks,colorlinks,breaklinks=true,urlcolor=color2,citecolor=color1,linkcolor=color1,bookmarksopen=false,pdftitle={Title},pdfauthor={Author}}
\usepackage{graphicx}
\JournalInfo{arXiv}

\Archive{}

\PaperTitle{Beam test results for the upgraded LHCb RICH optoelectronic readout system}

\Authors{P. Carniti\textsuperscript{1}* on behalf of the LHCb RICH collaboration}
\affiliation{\textsuperscript{1}\textit{INFN \& University of Milano Bicocca, Dip. di Fisica G. Occhialini, p.za della Scienza 3, 20126 Milano (Italy)}}
\affiliation{*\textbf{Corresponding author}: paolo.carniti@mib.infn.it} % Corresponding author

\newcommand{\keywordname}{Keywords}

\Abstract{Starting from 2018, the LHCb detector will be upgraded to operate at higher luminosity and extend its potential for new discoveries.
The Ring Imaging Cherenkov
(RICH) detectors are one of the key components for particle
identification of the LHCb detector and the upgraded specifications will require a redesign of the optoelectronic readout chain.
In the present work, we describe the experimental setup and the results of the tests carried out with a particle beam to assess and validate the performance of the optoelectronic readout system.}

\begin{document}

\flushbottom
\maketitle

\section{Introduction}

The LHCb experiment~\cite{Alves:2008zz} is devoted to high-precision measurements of CP violation and to search for New Physics by studying the decays of beauty and charmed hadrons produced at the Large Hadron Collider (LHC).

Two Ring Imaging Cherenkov (RICH) detectors are currently installed and operating successfully, providing a crucial role in the particle identification system of the LHCb experiment.

Starting from 2018, the LHCb experiment will be upgraded to operate at higher luminosity, up to $\mathrm{2\times10^{33}\ cm^{-2}s^{-1}}$, extending its potential for discovery and study of new phenomena.

Both the RICH detectors will be upgraded~\cite{LHCb-TDR-014} and the entire optoelectronic system has been redesigned in order to cope with the higher readout rate and increased occupancy.

The new photodetectors, readout electronics, mechanical assembly and cooling system have reached the final phase of development and their performance was validated during several beam test sessions from 2014 to 2016.

\subsection{Elementary cell}

\begin{figure}
	\centering
	\includegraphics[width=0.99\linewidth]{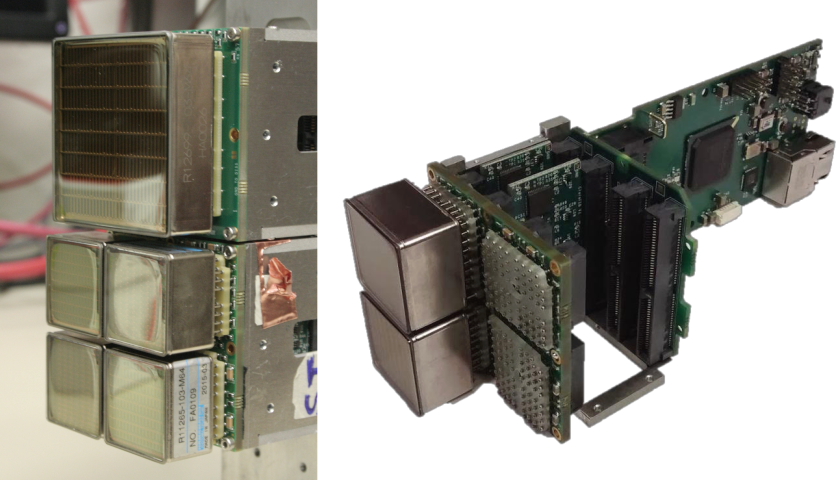}
	\caption{Photographs of the elementary cell (EC). The left photo shows the two different types of EC (EC-H at the top and EC-R at the bottom) with different sized MaPMTs.}
	\label{fig:ec}
\end{figure}

The photodetector planes will cover a total area of $\mathrm{3.7\ m^{2}}$ and will be installed in a tight space inside the magnetic shield box of the RICH detectors.
In order to maximise compactness and versatility of the detector, the optoelectronic readout system has been segmented in small and autonomous modules, called Elementary Cells (ECs), shown in Figure \ref{fig:ec}.

Two versions of the EC were built: EC-R with small $\mathrm{2.9\times2.9\ mm^{2}}$ pixels for the higher occupancy regions expected in RICH1 and in the inner part of RICH2, and EC-H with $\mathrm{6\times6\ mm^{2}}$ pixels for outer parts of RICH2.

The EC will use commercial Hamamatsu 64-pixels Multi-anode PMTs (MaPMTs) as photodetectors, read out by a custom radiation-hard ASIC for analogue-to-binary conversion, named CLARO~\cite{Carniti:2012ue,Andreotti:2015hga}.
Single pixel digital signals are then processed by an FPGA mounted on the Digital Board.

The MaPMT selected for EC-R is a customized version of the Hamamatsu R11265~\cite{Cadamuro:2014hza}, while EC-H will adopt a customized version of the R12699~\cite{calvi2015characterization}.
The MaPMTs will be installed on a custom baseboard which hosts the sockets and the HV bias resistor chain.
The CLARO chips will be hosted on front-end boards, installed between the baseboard and the backboard, which acts as a socket for the Digital Board.
The EC optoelectronic components are enclosed in a mechanical assembly which integrates with the cooling system.

\section{Experimental setup}

Beam tests were performed in several periods from 2014 to 2016 at the CERN SPS beam facility.
The beam used is composed mainly of protons and pions with a momentum of 180 GeV/c.

\begin{figure}
	\centering
	\includegraphics[width=0.45\linewidth,angle=90]{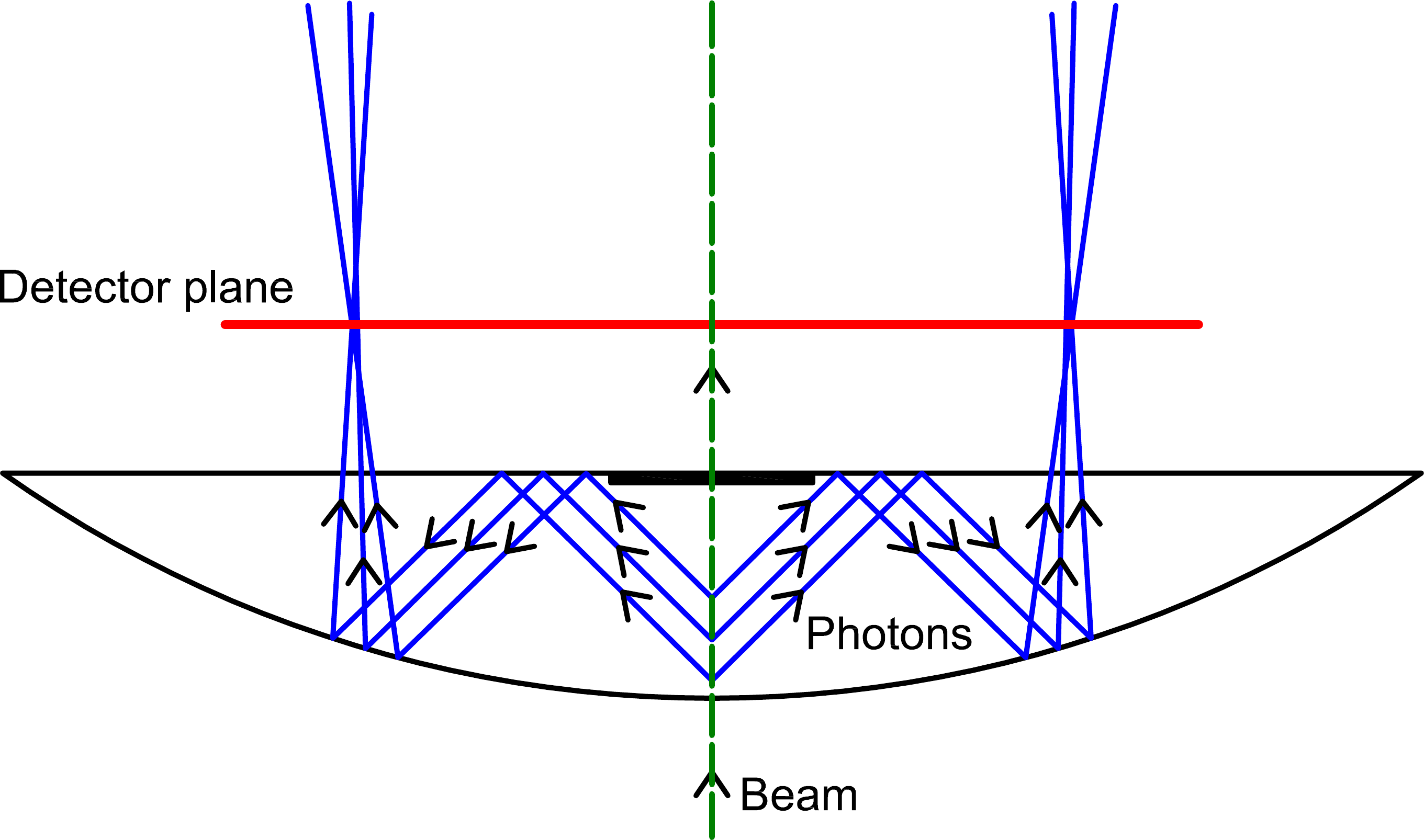}
	\caption{Schematic of the borosilicate glass lens used both as Cherenkov light radiator and focusing device.}
	\label{fig:lens}
\end{figure}

The beam enters a light-tight box and passes through a borosilicate glass lens which acts both as Cherenkov radiator and focusing device.
Figure \ref{fig:lens} shows the schematic of the optical setup.

A beam trigger is generated by two scintillator planes located upstream and downstream of the light-tight box.
A tracking telescope is also installed for recording and reconstructing particle tracks.
The system is controlled and monitored online with Java GUIs, WinCC\footnote{Siemens SIMATIC WinCC Open Architecture} control software and the ROOT analysis framework.

\begin{figure}
	\centering
	\includegraphics[width=0.65\linewidth]{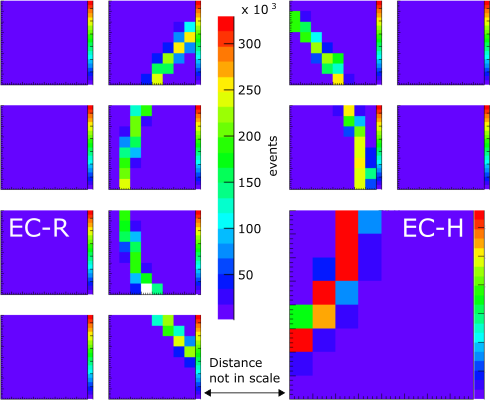}
	\caption{Accumulated Cherenkov rings on the online monitoring software.}
	\label{fig:ring}
\end{figure}

A typical display of accumulated Cherenkov rings, viewed on the online monitor software, is shown in Figure \ref{fig:ring}.

\section{Simulation}

The beam tests were preceded by comprehensive software simulations in order to optimise the setup and evaluate its performance.

\begin{figure}
	\centering
	\includegraphics[width=0.99\linewidth]{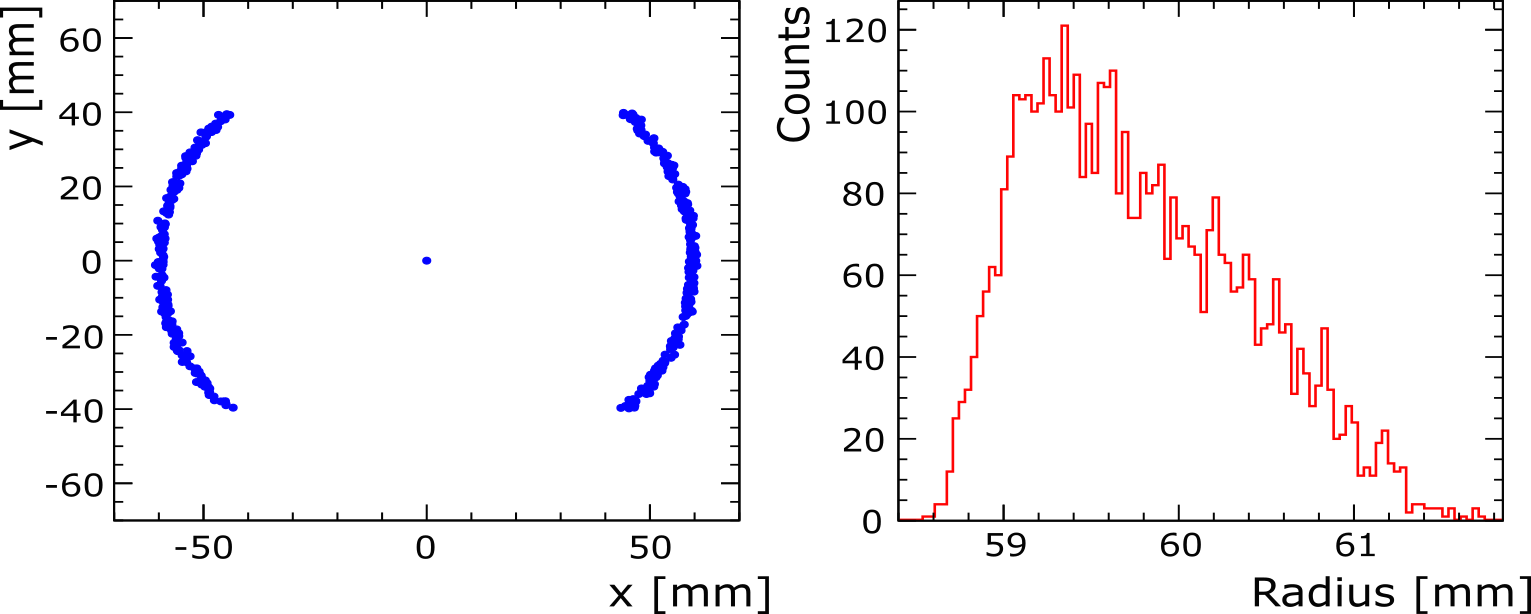}
	\caption{Optical simulations results. On the \textit{left} the expected ring shape. On the \textit{right} the radius distribution which gives an RMS of 0.6 mm.}
	\label{fig:simu}
\end{figure}

A ray-tracing simulation of the chosen lens configuration allowed the optimization of the system positioning and to calculate the expected ring radius and resolution, as shown in Figure \ref{fig:simu}.
In order to minimise the error due to the unknown emission point of Cherenkov photons in the lens, a black-tape mask was used to select only photons produced in the first 13 mm of the glass.

A more detailed simulation, using the \textsc{Geant4} toolkit, was performed on the complete system model.
This simulation used an accurate and complete representation of the test system, starting from the beam data recorded with the telescope, optical and geometrical specifications, measurements of the MaPMT's photocathode quantum efficiency, and also measurements of the electronic readout efficiency.
This simulation allowed the calculation of the expected number of detected photons per charged track and the expected Cherenkov angle.

\section{Results} 

The aim of these beam tests is to ensure that both performance and reliability meet the design specifications.
Another crucial aspect is the seamless integration of different subsystems.
In order to assess these goals, several measurements were performed during each beam test period.

%A first set of low-level measurements were dedicated to ensure that every single hardware component is working as expected in the beam test setup. Soon after, since the system proved to work extremely well even from the very first beam test, high-level measurements were performed and compared with simulation models.
Low-level measurements ensured that all hardware components work as expected in the beam setup. Then, high-level measurements were performed and the results compared with the simulation.
In the following paragraphs, the main results will be summarized.

\begin{figure}
	\centering
	\includegraphics[width=0.6\linewidth]{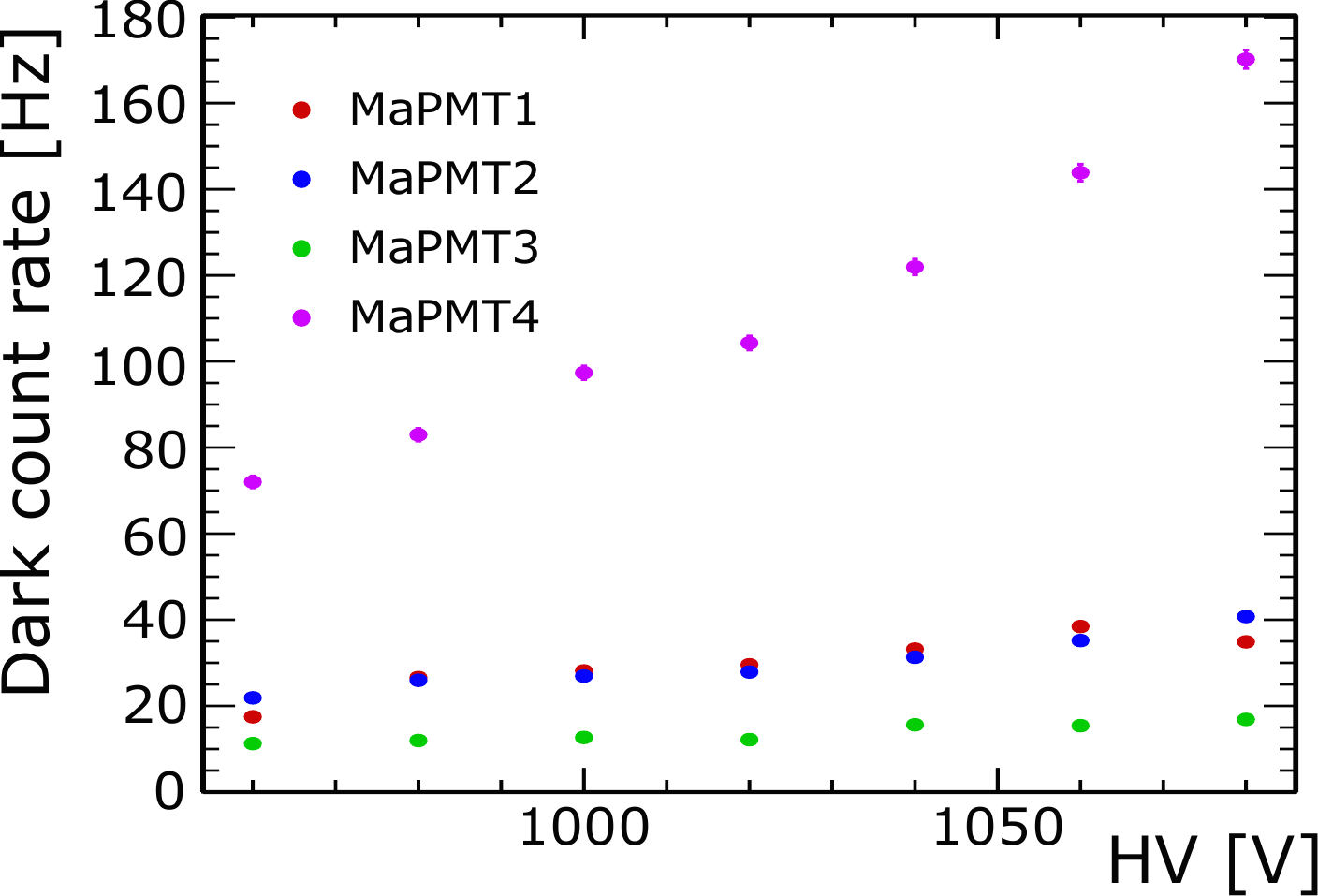}
	\caption{Dark count rates vs bias voltage for different MaPMTs.}
	\label{fig:dark_counts}
\end{figure}

\paragraph{Dark count rates}
Dark count rates were measured with different MaPMT HV settings. Typical rates spanned from a few tens of Hz, up to 160 Hz. An approximately linear dependence on bias voltage over a limited range was observed, as seen in Figure \ref{fig:dark_counts}.

\begin{figure}
	\centering
	\includegraphics[width=0.77\linewidth]{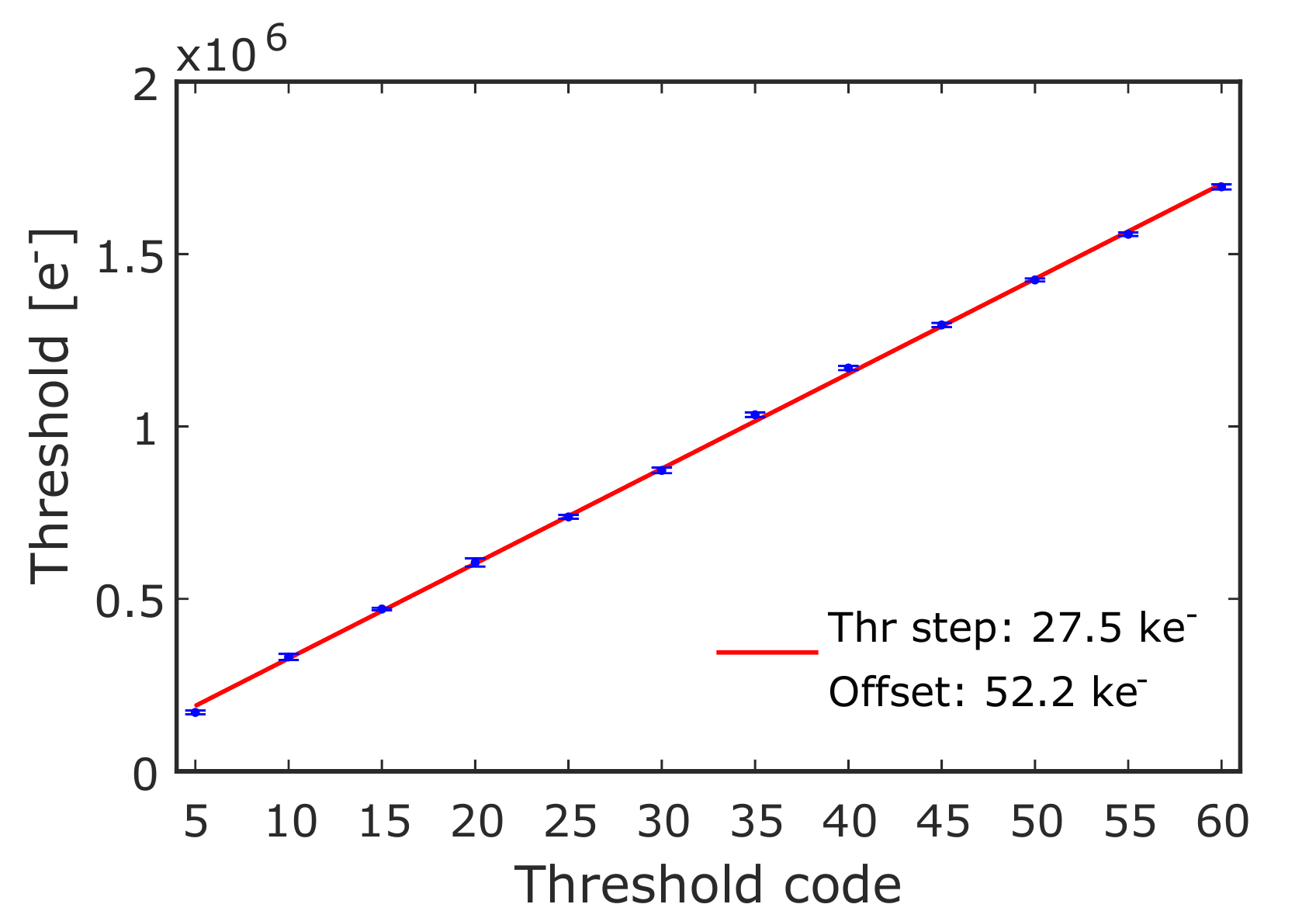}
	\caption{Threshold linearity for a typical channel, combining the results from several DAC scans performed at different CLARO threshold settings.}
	\label{fig:thr}
\end{figure}

\paragraph{CLARO threshold calibration}
The CLARO thresholds were calibrated using a DAC mounted on the backboard and a test capacitor incorporated in the ASIC.
The measurement was repeated at different thresholds to evaluate threshold linearity and extrapolated to determine the offset at threshold 0.
The results for a typical pixel are shown in Figure \ref{fig:thr}.

\begin{figure}
	\centering
	\includegraphics[width=0.7\linewidth]{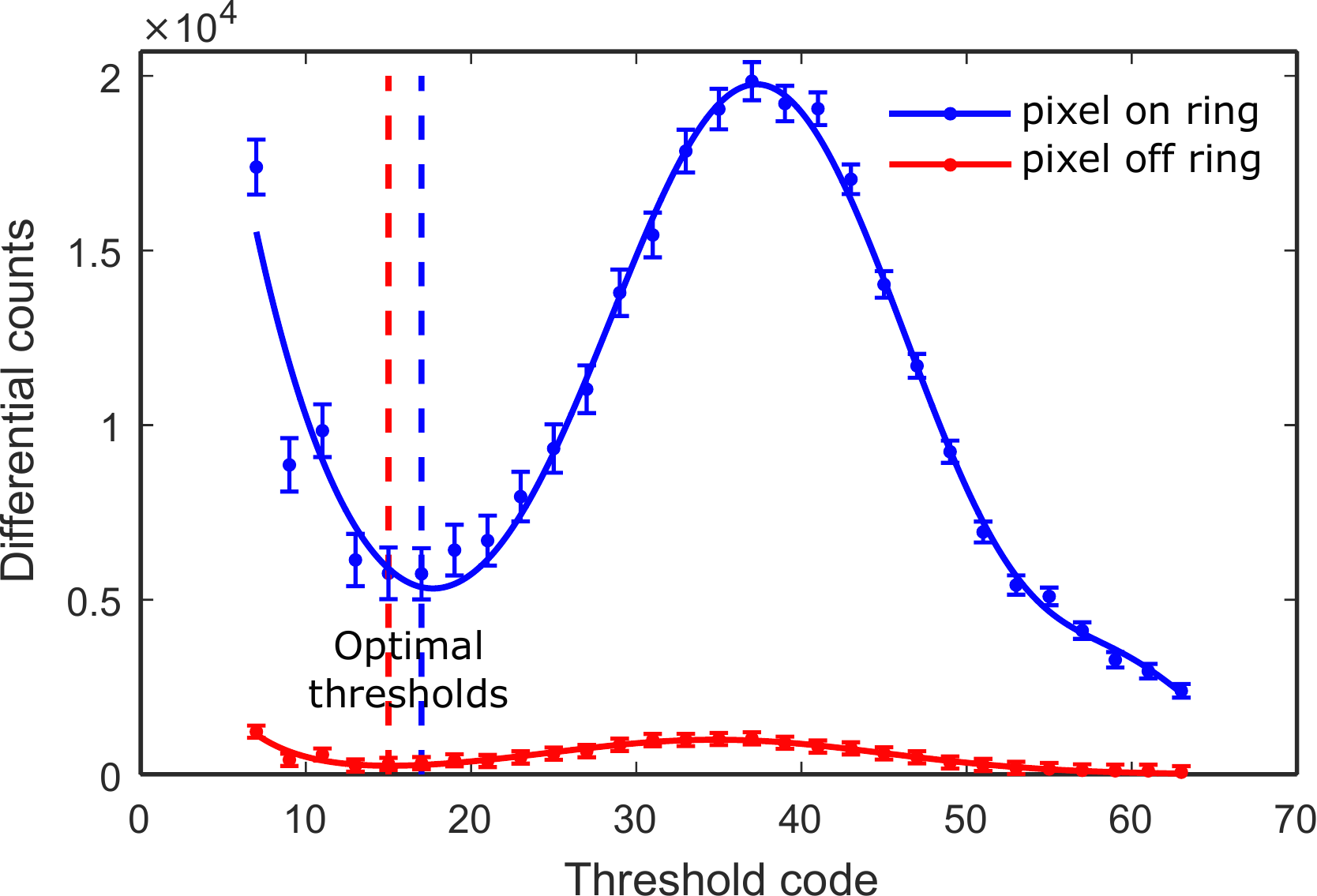}
	\caption{Single photon spectra for typical pixels. Dashed lines highlight the thresholds which are set during data acquisition.}
	\label{fig:spectra}
\end{figure}

\paragraph{Single photon peak reconstruction} 
Threshold scans were differentiated to reconstruct the single photon spectra on each pixel.
Figure \ref{fig:spectra} shows typical spectra for two pixels, one on the Cherenkov ring and one off the ring, where illumination is much lower and due to stray photons.
However, the peak for both pixels is very well resolved. 
This measurement allowed a consistent choice of the thresholds used in data acquisition for each pixel, placing it in the valley between the noise pedestal and the single photon peak, thus ensuring good noise rejection while maintaining a high detection efficiency.

\paragraph{Thermal measurements}
Several thermometers were installed inside the EC to monitor the temperature of each different component.
Tests were performed by changing the temperature of the cooling system and switching on and off different devices.
No critical issues were found and the cooling system proved satisfactory.

\begin{figure}
	\centering
	\includegraphics[width=0.8\linewidth]{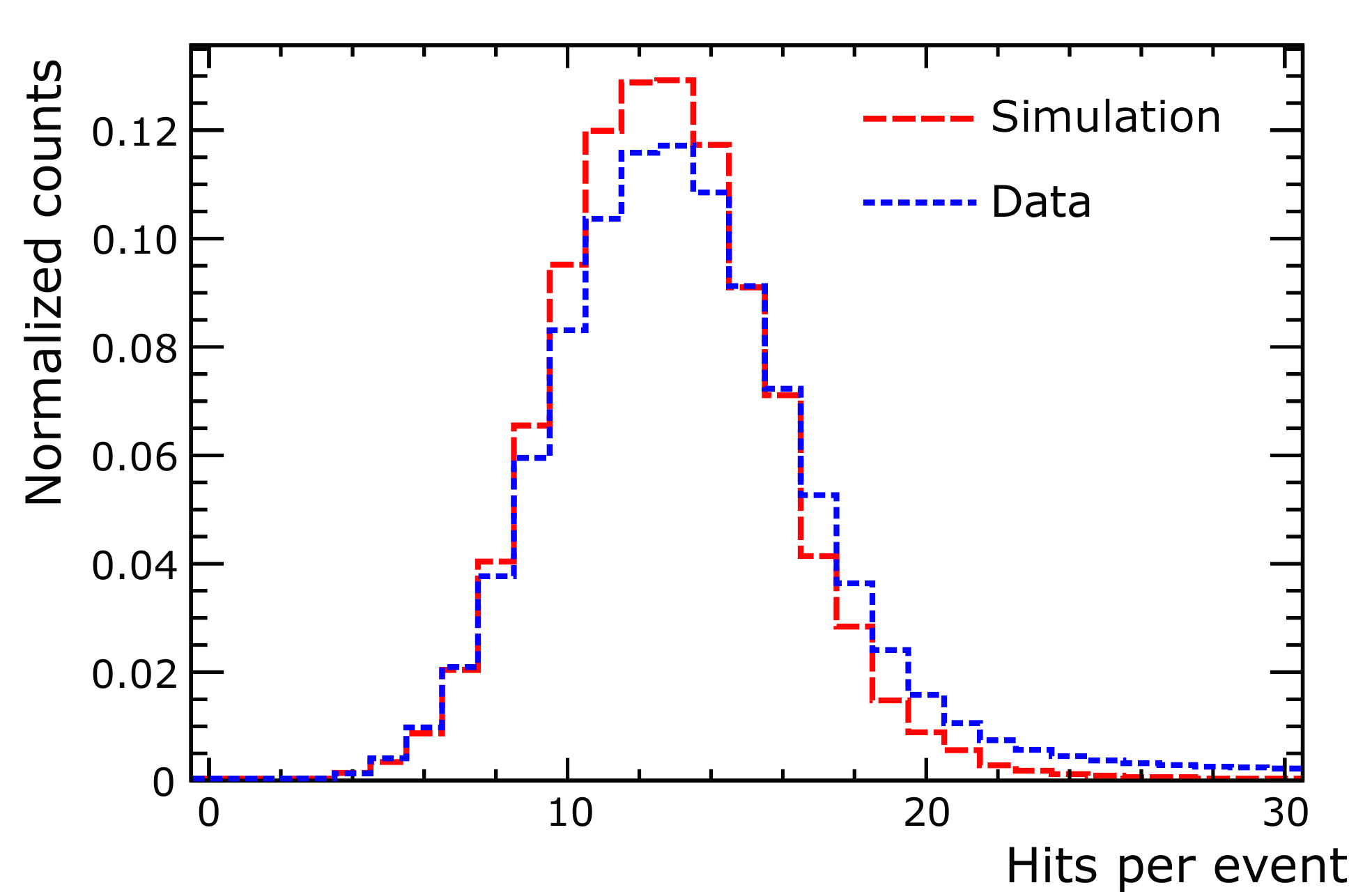}
	\caption{Distribution of hit multiplicity per event in simulated and in measured data.}
	\label{fig:n_hits}
\end{figure}

\paragraph{Photo-electron yield and multi-track correlation studies}
Figure \ref{fig:n_hits} shows the distribution of hits per event for both real data and \textsc{Geant4} simulation.
Using the information from the telescope, the behaviour of the system with respect to the number of incoming tracks was also studied.
The expected correlation was observed.

\begin{figure}
	\centering
	\includegraphics[width=0.75\linewidth]{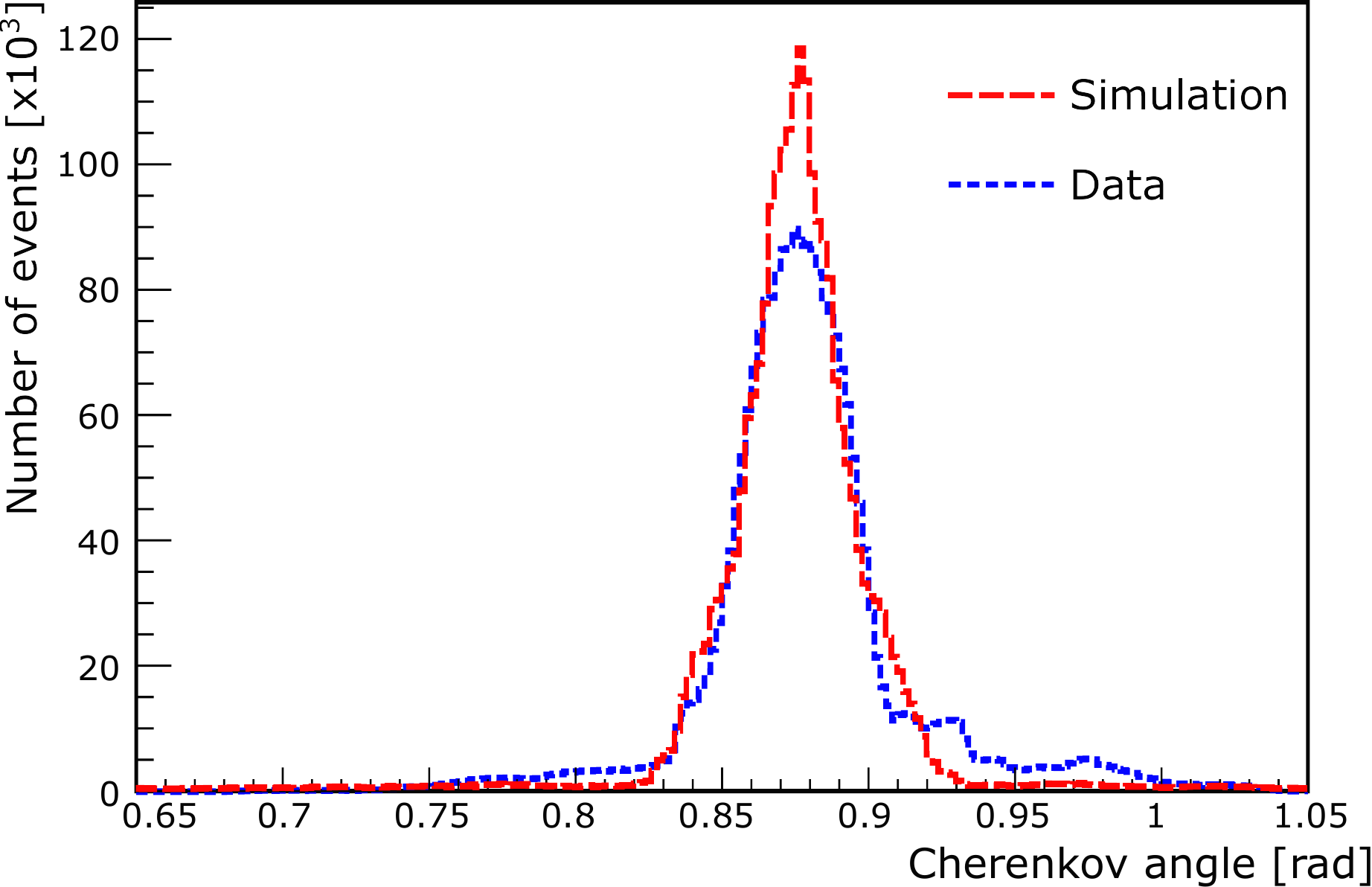}
	\caption{Measured and simulated Cherenkov angle distributions. The main peak for both real and simulated data is centred at 875 mrad. Sigma is 17 mrad.}
	\label{fig:ch_ang}
\end{figure}

\paragraph{Cherenkov angle reconstruction}
By fitting the ring event-by-event, it was possible to reconstruct the Cherenkov angle distribution. Figure \ref{fig:ch_ang} shows a comparison between real data and simulation. The measured angle is 875 mrad with a resolution of 17 mrad, dominated by the error due to the pixel size.

\section{Conclusions}

The beam test periods performed between 2014 and 2016 demonstrated the good performance of the optoelectronic readout system designed for the upgraded LHCb RICH with a realistic experimental environment.

Further beam tests are planned for the next months, using the final version of the Digital Board which will allow the readout rate to be increased up to the required 40 MHz.

\bibliography{mybibfile}

\end{document}